\documentclass[12pt]{article}
\usepackage{epsfig}

\usepackage{slashed}
\newcommand{\notp}{{\slashed{p}}}

\begin{document}

\title{
\vskip-3cm{\baselineskip14pt
\centerline{\normalsize DESY 06-141\hfill ISSN 0418-9833}
\centerline{\normalsize MPP-2006-108\hfill}
\centerline{\normalsize NYU-TH/06/08/29\hfill}
\centerline{\normalsize hep-ph/0608306\hfill}
\centerline{\normalsize August 2006\hfill}}
\vskip1.5cm
\bf Simple Approach to Renormalize the Cabibbo-Kobayashi-Maskawa Matrix}

\author{Bernd A. Kniehl\thanks{Electronic address: {\tt bernd.kniehl@desy.de};
permanent address:
II. Institut f\"ur Theoretische Physik, Universit\"at Hamburg,
Luruper Chaussee 149, 22761 Hamburg, Germany.}
\ and Alberto Sirlin\thanks{Electronic address: {\tt alberto.sirlin@nyu.edu};
permanent address: Department of Physics, New York University,
4 Washington Place, New York, New York 10003, USA.}\\
\\
{\normalsize\it Max-Planck-Institut f\"ur Physik
(Werner-Heisenberg-Institut),}\\
{\normalsize\it F\"ohringer Ring 6, 80805 Munich, Germany}}

\date{}

\maketitle

\begin{abstract}
We present an on-shell scheme to renormalize the Cabibbo-Kobayashi-Maskawa
(CKM) matrix.
It is based on a novel procedure to separate the external-leg mixing
corrections into gauge-independent self-mass and gauge-dependent wave-function
renormalization contributions, and to implement the on-shell renormalization
of the former with non-diagonal mass counterterm matrices.
Diagonalization of the complete mass matrix leads to an explicit CKM
counterterm matrix, which automatically satisfies all the following important
properties: it is gauge independent, preserves unitarity, and leads to
renormalized amplitudes that are non-singular in the limit in which any two
fermions become mass degenerate.

\medskip

\noindent
PACS: 11.10.Gh, 12.15.Ff, 12.15.Lk, 13.38.Be
\end{abstract}

\newpage

The Cabibbo-Kobayashi-Maskawa (CKM) \cite{cab} flavor mixing matrix, which
rules the charged-current interactions of the quark mass eigenstates and
describes how the heavier ones decay to the lighter ones, is one of the
fundamental cornerstones of the Standard Model of elementary particle physics
and, in particular, it is the key to our understanding why the weak
interactions are not invariant under simultaneous charge-conjugation and
parity transformations.
In fact, the detailed determination of this matrix is one of the major aims of 
recent experiments carried out at the $B$ factories \cite{pdg}, as well as the
objective of a wide range of theoretical studies \cite{pdg,Czarnecki:2004cw}.
An important theoretical problem associated with the CKM matrix is its
renormalization.
An early discussion, in the two-generation framework, was given in
Ref.~\cite{Marciano:1975cn}, focusing mostly on the cancellation of ultraviolet
divergences.
More recently, there have been a number of interesting papers that address the
renormalization of both the divergent and finite contributions at various
levels of generality and complexity \cite{Denner:1990yz}.

\begin{figure}[ht]
\begin{center}
\includegraphics[bb=112 626 524 779,width=0.49\textwidth]{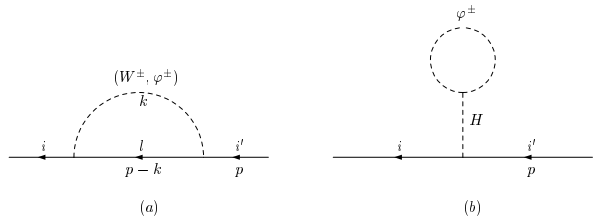}
\end{center}
\caption{\label{fig:one}%
Fermion mixing self-energy diagrams.
$H$ and $\phi^\pm$ denote Higgs and charged Goldstone bosons, respectively.
Diagram~(b) is included to cancel the gauge dependence in the diagonal
contribution of diagrams~(a).}
\end{figure}

In this Letter we propose an explicit on-shell framework to renormalize the
CKM matrix at the one-loop level, based on a novel procedure to separate the
external-leg mixing corrections into gauge-independent ``self-mass'' (sm) and
gauge-dependent ``wave-function renormalization'' (wfr) contributions, and to
implement the on-shell renormalization of the former with non-diagonal mass
counterterm matrices.
This procedure may be regarded as a simple generalization of Feynman's
approach in Quantum Electrodynamics (QED) \cite{Feynman:1949zx}.
We recall that, in QED, the self-energy contribution to an outgoing fermion is
given by
\begin{eqnarray}
\Delta{\cal M}^{\rm leg}&=&\overline{u}(p)\Sigma(\notp)\frac{1}{\notp-m},
\label{eq:dm}\\
\Sigma(\notp)&=&A+B(\notp-m)+\Sigma_{\rm fin}(\notp),
\label{eq:sig}
\end{eqnarray}
where $\Sigma(\notp)$ is the self-energy, $A$ and $B$ are divergent constants,
and $\Sigma_{\rm fin}(\notp)$ is a finite part which is proportional to
$(\notp-m)^2$ in the vicinity of $\notp=m$ and, therefore, vanishes when
inserted in Eq.~(\ref{eq:dm}).
The contribution of $A$ to Eq.~(\ref{eq:dm}) exhibits a pole at $\notp=m$ and
is gauge independent, while that of $B$ is regular at this point, but gauge
dependent.
They are referred to as sm and wfr contributions, respectively.
$A$ is canceled by the mass counterterm.
On the other hand, since the factor $(\notp-m)$ cancels the propagator's
singularity, in Feynman's approach $B$ is combined with the proper vertex
diagrams leading to a gauge-independent result.

In the case of the CKM matrix, one encounters not only diagonal terms as in
Eq.~(\ref{eq:dm}), but also off-diagonal external-leg contributions generated
by the Feynman diagrams of Fig.~\ref{fig:one}(a).
As a consequence, the self-energy corrections to an external leg are of the
form
\begin{equation}
\Delta{\cal M}_{ii^\prime}^{\rm leg}=\overline{u}_i(p)\Sigma_{ii^\prime}(\notp)
\frac{1}{\notp-m_{i^\prime}},
\label{eq:dmii}
\end{equation}
where $i$ denotes the external quark of momentum $p$ and mass $m_i$, and
$i^\prime$ the virtual quark of mass $m_{i^\prime}$.

We evaluate the contributions of Fig.~\ref{fig:one} in $R_\xi$ gauge, treating
the $i$ and $i^\prime$ quarks on an equal footing.
(A detailed account of our analytical work will be presented in a later, longer
manuscript \cite{long}.)
For example, we write
\begin{eqnarray}
2\notp a_-&=&\notp a_-+a_+\notp
\\
&=&(\notp-m_i)a_-+a_+(\notp-m_{i^\prime})+m_ia_-+m_{i^\prime}a_+,
\nonumber
\end{eqnarray}
where $a_\pm=(1\pm\gamma_5)/2$ are the chiral projectors.
Using this approach, we find that the contributions of Fig.~\ref{fig:one} can
be classified in four classes:
(i) terms with a left factor $(\notp-m_i)$;
(ii) terms with a right factor $(\notp-m_{i^\prime})$;
(iii) terms with a left factor $(\notp-m_i)$ and a right factor
$(\notp-m_{i^\prime})$; and
(iv) constant terms not involving $\notp$.
When inserted in Eq.~(\ref{eq:dmii}), the terms of class (iii) obviously
vanish, in analogy with $\Sigma_{\rm fin}(\notp)$ in Eqs.~(\ref{eq:dm}) and
(\ref{eq:sig}).
The terms of classes (i) and (ii) contain gauge-dependent parts but, when
inserted in Eq.~(\ref{eq:dmii}), they combine to cancel the propagator
$(\notp-m_{i^\prime})^{-1}$ in both the diagonal ($i=i^\prime$) and
off-diagonal ($i\ne i^\prime$) contributions.
Thus, they lead to expressions suitable for combination with the proper vertex
diagrams.
In analogy with $B$ in Eqs.~(\ref{eq:dm}) and (\ref{eq:sig}), such expressions
are identified as wfr contributions.
They satisfy the following important property:
all the gauge-dependent and all the divergent wfr contributions to the basic
$W\to q_i+\overline{q}_j$ amplitude are independent of $i^\prime$.
Using the unitarity relation
$V_{il}V_{li^\prime}^\dagger V_{i^\prime j}=V_{il}\delta_{lj}$ (since the
cofactor of this expression depends on $m_l$, the summation over $l$ is
performed later), one then finds that the gauge-dependent and the divergent
wfr contributions to the $W\to q_i+\overline{q}_j$ amplitude are independent
of CKM matrix elements, except for an overall factor $V_{ij}$, and depend only
on the external-quark masses $m_i$ and $m_j$.
Since the one-loop proper vertex diagrams also only depend on $m_i$, $m_j$,
and an overall factor $V_{ij}$, this observation implies that the proof of
gauge independence and finiteness of the remaining one-loop corrections to the
$W\to q_i+\overline{q}_j$ amplitude is the same as in the unmixed,
single-generation case!

In contrast to the contributions of classes (i) and (ii) to
Eq.~(\ref{eq:dmii}), those of class (iv) lead to a multiple of
$(\notp-m_{i^\prime})^{-1}$ with a cofactor that involves $a_\pm$, but is
independent of $\notp$.
Thus, they are unsuitable to be combined with the proper vertex diagrams and
are expected to be separately gauge independent, as we indeed find.
In analogy with $A$ in Eqs.~(\ref{eq:dm}) and (\ref{eq:sig}), they are
identified with sm contributions.
Specifically, in the case of an outgoing up-type quark, the sm contributions
from Fig.~\ref{fig:one} are given by the gauge-independent expression
\begin{eqnarray}
\Delta{\cal M}_{ii^\prime}^{\rm sm}&=&
\frac{g^2}{32\pi^2}V_{il}V_{li^\prime}^\dagger
\overline{u}_i(p)\left\{m_i\left(1+\frac{m_i^2}{2m_W^2}\Delta\right)
\right.
\nonumber\\
&&{}+\left[m_ia_-+m_{i^\prime}a_+
+\frac{m_im_{i^\prime}}{2m_W^2}(m_ia_++m_{i^\prime}a_-)\right]
\nonumber\\
&&{}\times
\left[I\left(m_i^2,m_l\right)-J\left(m_i^2,m_l\right)\right]
\nonumber\\
&&{}-\frac{m_l^2}{2m_W^2}(m_ia_-+m_{i^\prime}a_+)
\left[3\Delta+I\left(m_i^2,m_l\right)
\right.
\nonumber\\
&&{}+\left.\left.J\left(m_i^2,m_l\right)\right]
\vphantom{\frac{m_i^2}{2m_W^2}}
\right\}
\frac{1}{\notp-m_{i^\prime}},
\label{eq:legsm}
\end{eqnarray}
where $g$ is the SU(2) gauge coupling,
$\Delta=1/(n-4)+[\gamma_E-\ln(4\pi)]/2+\ln(m_W/\mu)$,
$n$ is the space-time dimension,
$\mu$ is the 't~Hooft mass,
$\gamma_E$ is Euler's constant,
\begin{eqnarray}
\lefteqn{\{I(p^2,m_l);J(p^2,m_l)\}
=\int_0^1dx\,\{1;x\}}
\nonumber\\
&&{}\times
\ln\frac{m_l^2x+m_W^2(1-x)-p^2x(1-x)-i\varepsilon}{m_W^2},
\end{eqnarray}
and $m_l$ are the masses of the virtual down-type quarks in
Fig.~\ref{fig:one}(a).
Terms independent of $m_l$ within the curly brackets of Eq.~(\ref{eq:legsm})
lead to diagonal contributions on account of 
$V_{il}V_{li^\prime}^\dagger=\delta_{ii^\prime}$.
There are other sm contributions involving virtual $Z^0$, $\phi^0$, $\gamma$,
and $H$ bosons, as well as additional tadpole diagrams, but these are again
diagonal expressions of the usual kind.

In order to generate mass counterterms, we proceed as follows.
In the weak-eigenstate basis, the bare mass terms are of the form
$-\overline{\psi}_R^{\prime Q}m_0^{\prime Q}\psi_L^{\prime Q}+\mbox{h.c.}$,
where $\psi_L^{\prime Q}$ and $\psi_R^{\prime Q}$ are left- and right-handed
column spinors involving the three up-type ($Q=U$) and down-type ($Q=D$)
quarks, and $m_0^{\prime Q}$ are non-diagonal matrices.
Writing $m_0^{\prime Q}=m^{\prime Q}-\delta m^{\prime Q}$, where
$m^{\prime Q}$ and $\delta m^{\prime Q}$ are the renormalized and counterterm
mass matrices, we consider a biunitary transformation of the quark fields that
diagonalizes $m^{\prime Q}$ leading to diagonal and real renormalized mass
matrices $m^Q$ and to new non-diagonal mass counterterm matrices $\delta m^Q$.
In the new framework, the mass term is given by
\begin{eqnarray}
\lefteqn{-\overline{\psi}\left(m-\delta m^{(-)}a_--\delta m^{(+)}a_+\right)
\psi}
\nonumber\\
&=&-\overline{\psi}_R\left(m-\delta m^{(-)}\right)\psi_L
-\overline{\psi}_L\left(m-\delta m^{(+)}\right)\psi_R,\quad
\label{eq:mass}
\end{eqnarray}
where $m$ is real, diagonal, and positive, and $\delta m^{(-)}$ and
$\delta m^{(+)}$ are arbitrary non-diagonal matrices subject to the
hermiticity constraint
\begin{equation}
\delta m^{(+)}=\delta m^{(-)\dagger}.
\label{eq:her}
\end{equation}
Here we have not exhibited the superscript $Q$, but it is understood that $m$
and $\delta m^{(\pm)}$ stand for two different sets of matrices involving the
up- and down-type quarks.
As usual, the mass counterterms are included in the interaction Lagrangian.
Their contribution to the external-leg corrections is given by
$-\overline{u}_i(p)\left(\delta m_{ii^\prime}^{(-)}a_-
+\delta m_{ii^\prime}^{(+)}a_+\right)/$\break $(\notp-m_{i^\prime})$.
Next we adjust $\delta m_{ii^\prime}^{(\pm)}$ to cancel, as much as possible,
the sm contributions given in Eq.~(\ref{eq:legsm}).
The cancellation of the divergent parts is achieved by choosing
\begin{eqnarray}
\left(\delta m_{\rm div}^{(-)}\right)_{ii^\prime}&=&
\frac{g^2m_i}{64\pi^2m_W^2}\Delta
\left(\delta_{ii^\prime}m_i^2-3V_{il}V_{li^\prime}^\dagger m_l^2\right),
\nonumber\\
\left(\delta m_{\rm div}^{(+)}\right)_{ii^\prime}&=&
\frac{g^2m_{i^\prime}}{64\pi^2m_W^2}\Delta
\left(\delta_{ii^\prime}m_i^2-3V_{il}V_{li^\prime}^\dagger m_l^2\right),\quad
\label{eq:div}
\end{eqnarray}
which satisfies the hermiticity constraint of Eq.~(\ref{eq:her}).
Because the functions $I(p^2,m_l)$ and $J(p^2,m_l)$ are evaluated at
$p^2=m_i^2$ in the $ii^\prime$ channel (where $i$ and $i^\prime$ are the
external and virtual quarks, respectively) and at $p^2=m_{i^\prime}^2$ in the
$i^\prime i$ channel (where $i^\prime$ and $i$ are the external and virtual
quarks, respectively), it is easy to see that it is not possible to cancel all
the finite pieces of Eq.~(\ref{eq:legsm}) in all channels without
contradicting Eq.~(\ref{eq:her}).
In particular, we note that once the $\delta m_{ii^\prime}^{(\pm)}$ are chosen,
the $\delta m_{i^\prime i}^{(\pm)}$ are fixed by Eq.~(\ref{eq:her}).
For this reason, we employ the following renormalization prescription:
the mass counterterms are chosen to exactly cancel all the contributions to
Eq.~(\ref{eq:legsm}) in the $i^\prime=i$, $uc$, $ut$, and $ct$ channels, and
all the sm contributions in the $j^\prime=j$, $sd$, $bd$, and $bs$ channels in
the corresponding down-type-quark expression.
(Here $j$ and $j^\prime$ are the incoming and virtual down-type quarks,
respectively.)
This implies that, after mass renormalization, there are residual sm
contributions in the $cu$, $tu$, $tc$, $ds$, $db$, and $sb$ channels.
However, these residual contributions are finite, gauge independent, and
numerically very small.
In fact, the fractional corrections they induce in the real parts of $V_{ij}$
reach a maximum value of ${\cal O}(4\times10^{-6})$ for $V_{ts}$, and they are
much smaller in the case of several other CKM matrix elements.  
Since they are regular in the limits $m_{i^\prime}\to m_i$ or
$m_{j^\prime}\to m_j$, they may be regarded as additional finite and
gauge-independent contributions to wave-function renormalization that
happen to be very small.

We emphasize that with this renormalization prescription the sm corrections
are fully canceled in all channels in which the external particle is a $u$,
$\overline{u}$, $d$, or $\overline{d}$ quark.
This is of particular interest since $V_{ud}$, the parameter associated with
$W\to u+\overline{d}$, is by far the most precisely determined CKM matrix
element \cite{Czarnecki:2004cw}.

It is also interesting to note that, since Eq.~(\ref{eq:div}) satisfies
Eq.~(\ref{eq:her}), the modified minimal-subtraction
($\overline{\mathrm{MS}}$) renormalization, in which only the
$1/(n-4)+[\gamma_E-\ln(4\pi)]/2$ terms are subtracted, can be implemented in
all non-diagonal channels.
More generally, one can consider a renormalization prescription that satisfies
the hermiticity condition in all channels by choosing the mass counterterms to
cancel the off-diagonal terms in Eq.~(\ref{eq:legsm}) and the corresponding
down-type-quark expression with the functions $I(p^2,m_l)$ and $J(p^2,m_l)$
evaluated at the same fixed $p^2$ value for all flavors.
Since Eq.~(\ref{eq:legsm}) is explicitly gauge independent, in our formulation
there is no restriction in the choice of $p^2$ other than that it should not
generate imaginary parts in the integrals $I(p^2,m_l)$ and $J(p^2,m_l)$.
In particular, $p^2$ can have any value $p^2\le m_W^2$.
Of course, since it is desirable to cancel the sm contributions as much as
possible, it is convenient to choose $0\le p^2\ll m_W^2$.
It should be pointed out, however, that the $\overline{\mathrm{MS}}$ and
fixed-$p^2$ subtraction prescriptions of mass renormalization are not on-shell
schemes and lead to residual sm contributions in all off-diagonal channels,
which diverge in the limits $m_{i^\prime}\to m_i$ or $m_{j^\prime}\to m_j$.

An alternative formulation, equivalent to the one discussed so far, is
obtained by diagonalizing the complete mass matrix
$m-\delta m^{(-)}a_--\delta m^{(+)}a_+$ in Eq.~(\ref{eq:mass}).
This is achieved by a biunitary transformation
\begin{equation}
\psi_L=U_L\hat\psi_L,\qquad\psi_R=U_R\hat\psi_R.
\end{equation}
At the one-loop level, it is sufficient to approximate
\begin{equation} 
U_L=1+ih_L,\qquad U_R=1+ih_R,
\end{equation}
where $h_L$ and $h_R$ are hermitian matrices of ${\cal O}(g^2)$.
The diagonalization is implemented by choosing
\begin{equation}
i(h_L)_{ii^\prime}=\frac{m_i\delta m_{ii^\prime}^{(-)}
+\delta m_{ii^\prime}^{(+)}m_{i^\prime}}{m_i^2-m_{i^\prime}^2}
\qquad (i\ne i^\prime),
\label{eq:hlii}
\end{equation}
while $i(h_R)_{ii^\prime}$ is obtained by exchanging
$\delta m^{(-)}\leftrightarrow\delta m^{(+)}$ in Eq.~(\ref{eq:hlii}).
Since the only effect of the diagonal terms of $h_L$ and $h_R$ on the
$Wq_i\overline{q}_j$ interaction is to introduce phases that can be absorbed
in a redefinition of the quark fields, it is convenient to set
$(h_L)_{ii}=(h_R)_{ii}=0$.
This analysis is carried out separately to diagonalize the mass matrices
of the up- and down-type quarks.
Thus, we obtain two pairs of matrices: $h_L^U$ and $h_R^U$ for
the up-type quarks and $h_L^D$ and $h_R^D$ for the down-type quarks.
Next we consider the effect of this biunitary transformation on the
$Wq_i\overline{q}_j$ interaction
\begin{equation}
{\cal L}_{Wq_i\overline{q}_j}=-\frac{g_0}{\sqrt2}
\overline{\psi}_L^UV\gamma^\lambda\psi_L^DW_\lambda+\mbox{h.c.}.
\end{equation}
We readily find that
\begin{equation}
{\cal L}_{Wq_i\overline{q}_j}=-\frac{g_0}{\sqrt2}\overline{\hat\psi}_L^U
(V-\delta V)\gamma^\lambda\hat\psi_L^DW_\lambda+\mbox{h.c.},
\label{eq:hc}
\end{equation}
where
\begin{equation}
\delta V=i\left(h_L^UV-Vh_L^D\right).
\label{eq:dv}
\end{equation}
It is important to note that $V-\delta V$ satisfies the unitarity condition
through ${\cal O}(g^2)$:
\begin{equation}
(V-\delta V)^\dagger(V-\delta V)=1+{\cal O}(g^4).
\end{equation}
In the $(\hat\psi_L,\hat\psi_R)$ basis, in which the complete quark mass
matrices are diagonal, $\delta V$ and $V_0=V-\delta V$ represent the
counterterm and bare CKM matrices, respectively.
One readily verifies that the term $ih_L^UV$ in $\delta V$ leads to the same
off-diagonal contribution to the $W\to q_i+\overline{q}_j$ amplitude as
$\delta m^{U(-)}$ and $\delta m^{U(+)}$ in our previous discussion in the
$(\psi_L,\psi_R)$ basis.
Similarly, the term $-iVh_L^D$ leads to the same contributions as
$\delta m^{D(-)}$ and $\delta m^{D(+)}$.
It is important to emphasize that this formulation is consistent with the
unitarity and gauge independence of both the renormalized and bare CKM
matrices, $V$ and $V_0$, respectively.

For completeness, we exhibit the CKM counterterm matrix in component form:
\begin{eqnarray}
\delta V_{ij}&=&i\left[\left(h_L^U\right)_{ii^\prime}V_{i^\prime j}
-V_{ij^\prime}\left(h_L^D\right)_{j^\prime j}\right]
\nonumber\\
&=&\frac{m_i^U\delta m_{ii^\prime}^{U(-)}
+\delta m_{ii^\prime}^{U(+)}m_{i^\prime}^U}{\left(m_i^U\right)^2
-\left(m_{i^\prime}^U\right)^2}V_{i^\prime j}
\nonumber\\
&&{}-V_{ij^\prime}\frac{m_{j^\prime}^D\delta m_{j^\prime j}^{D(-)}
+\delta m_{j^\prime j}^{D(+)}m_j^D}{\left(m_{j^\prime}^D\right)^2
-\left(m_j^D\right)^2},
\label{eq:dvii}
\end{eqnarray}
where it is understood that $i^\prime\ne i$ in the first term on the
r.h.s.\ and $j^\prime\ne j$ in the second, and $\delta m_{ii^\prime}^{U(\pm)}$
and $\delta m_{j^\prime j}^{D(\pm)}$ are the off-diagonal mass counterterms
determined by the on-shell renormalization prescriptions proposed in our first
formulation.
The coefficient of $1/(n-4)$ in Eq.~(\ref{eq:dvii}) is, of course, common to
all renormalization prescriptions for the CKM matrix \cite{Denner:1990yz} and
also appears in its renormalization group equation \cite{Babu:1987im}.

In summary, after introducing a novel procedure to separate the external-leg
mixing corrections into gauge-independent sm and gauge-dependent wfr
contributions, in analogy with Feynman's treatment in QED, we have implemented
their renormalization in two equivalent frameworks.
The first one is carried out in a basis in which the renormalized quark
matrices are diagonal and the non-diagonal mass counterterm matrices are
employed to cancel all the divergent sm contributions, and also their finite
parts up to hermiticity constraints.
In particular, the sm corrections are fully canceled in the
$W\to u+\overline{d}$ amplitude, associated with $V_{ud}$, the most accurately
measured CKM parameter.
Residual finite contributions in other channels are very small.
We have also pointed out that the proof of gauge independence and finiteness
of the remaining one-loop corrections to the $W\to q_i+\overline{q}_j$
amplitude reduces to that in the unmixed, single-generation case.
Alternative renormalization prescriptions that are ``democratic,'' in the sense
that they do not single out particular off-diagonal channels, were briefly
outlined.
However, strictly speaking, they are not on-shell schemes and lead to residual
sm contributions in all off-diagonal channels, which diverge in the limits
$m_{i^\prime}\to m_i$ or $m_{j^\prime}\to m_j$.

The second formulation was obtained by diagonalizing the complete mass
matrices, namely the renormalized plus counterterm mass matrices derived in
the first approach.
In the second framework a CKM counterterm matrix $\delta V$ was generated which
again cancels the divergent and, to the extent allowed by the hermiticity
constraints, also the finite parts of the off-diagonal sm contributions.
As usual, the diagonal sm contributions are canceled by the mass counterterms,
which in this approach are also diagonal.
An important feature is that this formulation is consistent with the unitarity
and gauge independence of both the renormalized and bare CKM matrices, $V$ and
$V_0=V-\delta V$, respectively.

As is well known, an enduring difficulty, thirty years old, in a satisfactory
treatment of the one-loop electroweak corrections to all charged-current
processes involving fermions is due to the external off-diagonal self-energy
effects depicted in Fig.~\ref{fig:one}(a).
Since the mass renormalization of the usual, diagonal effects must necessarily
involve a complete subtraction of the sm contributions to avoid the
propagator's singularity [see Eq.~(\ref{eq:dm})], it is natural to follow the
same strategy in the off-diagonal contributions.
Thus, an on-shell renormalization procedure to treat all these effects is
highly desirable and strongly motivated.
Such an objective has been achieved for the first time in this Letter in a way
that the following important properties are manifestly satisfied:
the CKM counterterm matrix is gauge independent, preserves unitarity, and leads
to renormalized amplitudes that are non-singular in the limit in which any two
fermions become mass degenerate.
Because of the close analogy with QED and the fact that our decomposition
procedure is algebraic in nature, it is likely that this approach can be
naturally generalized to higher orders.

We are grateful to the Max Planck Institute for Physics in Munich for the
hospitality during a visit when this manuscript was prepared.
The work of B.A.K. was supported in part by the German Research Foundation
through the Collaborative Research Center No.\ 676 {\it Particles, Strings and
the Early Universe---the Structure of Matter and Space-Time}.
The work of A.S. was supported in part by the Alexander von Humboldt
Foundation through the Humboldt Research Award No.\ IV~USA~1051120~USS and by
the National Science Foundation Grant No.\ PHY-0245068.

\end{document}